\documentclass[nofootinbib,twocolumn,preprintnumbers]{revtex4-1}
\pdfoutput=1
\usepackage{amsmath,amsthm,amssymb,multirow,psfrag}
\usepackage{epsfig}
\usepackage{color}
\usepackage{slashed}
\usepackage{hyperref}
\graphicspath{{./Figures/}}

\begin{document}

\def\lsim{\mathrel{\rlap{\lower4pt\hbox{\hskip1pt$\sim$}}
  \raise1pt\hbox{$<$}}}
\def\gsim{\mathrel{\rlap{\lower4pt\hbox{\hskip1pt$\sim$}}
  \raise1pt\hbox{$>$}}}
\newcommand{\vev}[1]{ \left\langle {#1} \right\rangle }
\newcommand{\bra}[1]{ \langle {#1} | }
\newcommand{\ket}[1]{ | {#1} \rangle }
\newcommand{\ev}{ {\rm eV} }
\newcommand{\kev}{{\rm keV}}
\newcommand{\mev}{{\rm MeV}}
\newcommand{\gev}{{\mathrm GeV}}
\newcommand{\tev}{{\rm TeV}}
\newcommand{\mpl}{$M_{Pl}$}
\newcommand{\mw}{$M_{W}$}
\newcommand{\Ft}{F_{T}}
\newcommand{\Zparity}{\mathbb{Z}_2}
\newcommand{\BLambda}{\boldsymbol{\lambda}}
\newcommand{\met}{\;\not\!\!\!{E}_T}
\newcommand{\beq}{\begin{equation}}
\newcommand{\eeq}{\end{equation}}
\newcommand{\bea}{\begin{eqnarray}}
\newcommand{\eea}{\end{eqnarray}}
\newcommand{\nn}{\nonumber}
\newcommand{\hc}{\mathrm{h.c.}}
\newcommand{\eps}{\epsilon}
\newcommand{\bwt}{\begin{widetext}}
\newcommand{\ewt}{\end{widetext}}
\newcommand{\draftnote}[1]{{\bf\color{blue} #1}}

\newcommand{\cO}{{\cal O}}
\newcommand{\cL}{{\cal L}}
\newcommand{\cM}{{\cal M}}

\newcommand{\fref}[1]{Fig.~\ref{fig:#1}} 
\newcommand{\eref}[1]{Eq.~\eqref{eq:#1}} 
\newcommand{\aref}[1]{Appendix~\ref{app:#1}}
\newcommand{\sref}[1]{Section~\ref{sec:#1}}
\newcommand{\tref}[1]{Table~\ref{tab:#1}}

\title{\LARGE{{\bf{Emerging Jets Displaced into the Future} }}}

\author{{\bf {Paul Archer-Smith, Dylan Linthorne, Daniel Stolarski}}}

\affiliation{
Ottawa-Carleton  Institute  for  Physics,  Carleton  University,\\
1125  Colonel  By  Drive,  Ottawa,  Ontario  K1S  5B6,  Canada
}

\email{
PaulSmith3@cmail.carleton.ca\\
Dylan.Linthorne@carleton.ca \\
Stolar@physics.carleton.ca
}

\begin{abstract}
We examine the potential of future long-lived particle experiments to probe dark QCD models that feature Emerging Jets. The core of this analysis focuses on the transverse detectors AL3X, ANUBIS, CODEX-b, and MATHUSLA as they cover the most relevant parameter space, though the highly forward experiments MAPP, FORMOSA, and FASER are also explored. Geometric coverage of the detectors is calculated and used to determine the number of signal events and kinematic distribution measured for a collection of different benchmark models. This is used to map out the discovery potential of the Emerging Jets parameter space. Although all experiments demonstrate some reach, AL3X, ANUBIS, and MATHUSLA stand out as the most promising for exploring the dark QCD Emerging Jets parameter space.  
\end{abstract}

\maketitle

\section{Introduction} 
\label{sec:intro} 


Long lived particles (LLPs), those that travel a macroscopic distance before decaying, arise in many well motivated new physics models including SUSY~\cite{Barbier:2004ez,Fan:2011yu,Arvanitaki:2012ps,Arkani-Hamed:2012fhg}, dark matter~\cite{Bai:2013xga,Lonsdale:2017mzg,Lonsdale:2018xwd}, and neutral naturalness~\cite{Craig:2015pha,Curtin:2015fna,Cheng:2016uqk,Kilic:2018sew}. While LLPs can be searched for at the LHC's current detectors~\cite{LLP2,LLP1}, a host of new detectors have been proposed in the last few years \cite{AL3X, ANUBIS,CODEXb,Gligorov:2017nwh,FASER,FASER2,FASER3,FORMOSA,GAZELLE,MAPP, MAPP12, MAPP2,MATHUSLA4,MATHUSLA,MATHUSLA2,MATHUSLA3,MilliQan,MilliQan2,SHiP1,SHiP2,SHiP3} that can open up new discovery windows for long lived particles. As of this writing, only FASER~\cite{FASER,FASER2,FASER3} has been approved, but R\&D is ongoing for all of these experiments, and future approval is a possibility for the rest. 


Particle phenomenologists have begun exploring the reach of a variety of these detectors for a wide class of models~\cite{deVries:2015mfw,Dercks:2018eua,Dercks:2018wum,Hirsch:2020klk,Dreiner:2020qbi,DeVries:2020jbs,Gehrlein:2021hsk,Cottin:2021lzz,Kamada:2021cow}, sometimes comparing the reach across different detectors. In this work we take a similar approach, comparing the discovery reach of a broad class of these detectors for Emerging Jets (EJ)~\cite{EJ}. EJ arise from a GeV scale confining dark sector with a TeV scale mediator, and are inspired by the asymmetric dark matter paradigm, and in particular the model of Bai and Schwaller~\cite{Bai:2013xga}. The EJ signature features clusters of LLPs produced in jet-like structures, and there has now been an experimental search for Emerging Jets at CMS~\cite{CMS:2018bvr}. There have also been phenomenological studies of the flavour effects in these models~\cite{Renner:2018fhh,Bensalem:2021qtj}, a recast of different bounds~\cite{EJPlus}, and a study of novel triggering strategies~\cite{EJTrig}.


In these models, the lifetime of the LLPs is given schematically by
\begin{equation}
c\tau \sim \frac{M^4}{g^4 m^5} \sim \frac{0.1 \, \text{m}}{g^4} \left(\frac{M}{\text{TeV}} \right)^4 \left(\frac{\text{GeV}}{m} \right)^5 \, ,
\end{equation}
where $m$ are mass scales associated with the confining hidden sector and the LLP mass, $M$ is the mediator mass, and $g$ is an unknown coupling does not necessarily control the production rate of Emerging Jets. A more detailed calculation of the lifetime is given in Eqs.~\eqref{eqn:DPdecay} and~\eqref{eqn:Lifetime}. Still, from this schematic equation we see that if $g \sim 0.1$, the lifetime of the LLPs can be significantly longer than the size of the LHC's current detectors, and the proposed detectors designed specifically to detect LLPs could be ideal discovery machines. A preliminary study of discovery potential for EJ at SHiP~\cite{SHiP2} and MATHUSLA~\cite{MATHUSLA4} has been undertaken, but here we seek to explore a much larger range of detectors as well as a broader parameter space of EJ models. 


In this work, we examine the discovery potential of Emerging Jets for eight different experiments: AL3X~\cite{AL3X}, ANUBIS~\cite{ANUBIS}, CODEX-b~\cite{CODEXb}, FASER~\cite{FASER}, FORMOSA~\cite{FORMOSA}, MAPP~\cite{MAPP}, MATHUSLA~\cite{MATHUSLA4}, and MilliQan~\cite{MilliQan} (other proposed LLP detectors, such as GAZELLE~\cite{GAZELLE} and SHiP~\cite{SHiP1,SHiP2,SHiP3}, that search for particles produced at beam lines with significantly lower energy are left for future work). Of these experiments, AL3X, ANUBIS, CODEX-b, and MATHUSLA cover the most relevant parameter space and are centred in this analysis. Geometric coverage of these detectors is calculated and used to determine the number of signal events and kinematic distributions measured for a collection of different model benchmarks. Sensitivity of these detectors to Emerging Jet models as a function of dark pion mass and lifetime is determined and the expected limits on the Emerging Jet parameter space are presented. Finally, the discovery reach for MAPP, FORMOSA, and FASER is also explored as these detectors are highly forward and thus sensitive to different production processes.  

The rest of the paper is organized as follows: Sec.~\ref{sec:jets} reviews the Emerging Jets framework and underlying BSM models used throughout the paper, Sec.~\ref{sec:detectors} provides an overview of LLP experiments, and Sec.~\ref{sec:events} discusses event generation and our simulation process.  Sec.~\ref{sec:results} presents our findings and Sec.~\ref{sec:conclusion} gives our conclusions. A discussion of forward detectors and the $t$-channel process is given in App.~\ref{sec:LLP}, and the public code used to simulate the detectors is briefly described in App.~\ref{sec:DS}.

\section{Emerging Jets}
\label{sec:jets}

The general model setup that we use to explore the detection prospects of Emerging Jets at LLP experiments is based on the original framework presented in \cite{EJ}. The SM gauge group is extended to include an additional \textit{dark QCD} symmetry:
\begin{equation}\label{eqn:gauge}
\mathrm{SU}(3)_C \times \mathrm{SU}(2)_L \times \mathrm{U}(1)_Y \times \mathrm{SU}(N_d).
\end{equation}
The first three terms are the SM gauge symmetries of colour, weak isospin, and hypercharge, respectively, while the final term introduces $N_d \geq 2$ dark colours. SM particles are, of course, singlets under this new group. $n_f$ new Dirac fermions --- \textit{dark quarks}, $Q_d$ --- are taken to exist within this model: these dark quarks are charged under dark QCD but remain singlets under the SM gauge groups.  The dark QCD sector confines at $\Lambda_d$ which leads to a spectrum of dark baryons and mesons clustered at this scale. Following in the footsteps of previous work \cite{EJ,EJTrig}, we take $\Lambda_d$ to be $\cO (\mathrm{5\,GeV})$; at this scale if there is an accidental dark baryon number symmetry (analogous to QCD) then the dark baryons could be dark matter. Further, this fits well with a picture of asymmetric dark matter~\cite{Bai:2013xga,Lonsdale:2017mzg,Lonsdale:2018xwd} as the energy density of dark matter is roughly five times that of visible matter: this makes this scale and its ability to generate dark matter at roughly $5\times m_{proton}$ especially appealing.


The dark QCD also features pseudo-Goldstone bosons (\textit{dark pions}, $\pi_d$) that play a similar role to that of the SM QCD pions. These dark pions are taken to have a common mass $m_{\pi_d} < \Lambda_d$, typically $m_{\pi_d} \sim 1 \,\mathrm{GeV}$. Unlike the dark baryons, there is no conserved \textit{dark meson number} and, as such, the dark pions can decay to SM particles. Heavier dark mesons have a lifetime of order $\Lambda_d^{-1}$ and can, if kinematically permitted, decay into dark pions (similar to the $\rho \rightarrow \pi\pi$ decay in the SM). For the parameter space considered within this paper, the dark pions are much lighter than the heavy dark hadrons and thus can be produced in much larger abundances than the stable dark baryons. 

In order to connect the SM and dark QCD sectors, a heavy mediator, $X_d$, is introduced (this resembles the hidden valley models of \cite{Strassler:2006im} and the dark QCD models of \cite{Bai:2013xga}). This complex scalar is a bifundamental under both SM and dark QCD; thus, $X_d$ can be pair produced at the LHC with each of these mediators decaying into an SM quark and a dark quark. Previous work on Emerging Jets also examined a vector mediator $Z^\prime$ \cite{EJ,EJTrig}, however since our goal is primarily to examine the relative parameter space coverage of various LLP detectors, we restrict ourselves to dealing with the scalar scenario.

The high-energy Lagrangian of the dark QCD parts of this theory is 
\begin{equation}\label{eqn:Lagrangian}
\begin{split}
\mathcal{L} \supset \bar{Q}_{d_i}(\slashed{D} - m_{d_i})Q_{d_i} + (D_\mu X_d)(D_\mu X_d)^\dagger - m^2_{X} X_d^2 \\
-\frac{1}{4}G^{\mu\nu}_d G_{d\mu\nu} + (\kappa_{ij}\bar{Q}_{d_i}q_j X_d + \mathrm{h.c.})
\end{split}
\end{equation} 
where $G^{\mu\nu}_d$ is the dark gluon field strength tensor, and the covariant derivatives contain all gauge field couplings. $m^2_X$ is the mass of the scalar mediator, and we take it to be large: $m_X \gg \Lambda_d$. The field $q$ is one of the SM quark multiplets, which for concreteness we take to be right handed down-type. This sets the electroweak charges of the $X_d$. Finally $\kappa_{ij}$ is an $n_f \times 3$ matrix containing the mediator's Yukawa couplings. Generically,  these type of Yukawa couplings could lead to large flavour violating processes which are further studied in~\cite{Renner:2018fhh,Bensalem:2021qtj}.

Broadly speaking, this type of model has everything necessary to produce Emerging Jets: a hierarchy is present between the masses of the mediator and dark sector hadrons, the dark sector is strongly coupled, and the dark sector particles feature macroscopic decay lengths. Explicitly, pair production of the mediators can occur at the LHC, which is followed by a decay of each $X_d$ to a quark and dark quark. The dark quarks shower and produce a large number of dark hadrons. The dark mesons ultimately decay back to SM particles, but this happens over macroscopic distances --- resulting in a dark jet that slowly emerges into a visible jet. 

With this in mind, we turn our attention back to the dark pions in order to flesh out the relevant details of the dark sector. The Yukawa couplings break the $n_f \times n_f$ dark flavour symmetry that the dark pions are the Goldstone bosons of. This gifts the dark pions their mass\footnote{The rank of the $\kappa$ matrix is at most 3, so  if $n_f > 3$, some pions are expected to remain massless.}, making decay to SM quarks kinematically accessible. Integrating out the mediator gives us the effective Lagrangian of the dark quarks:
\begin{equation}\label{eqn:EffLagrangian}
m_{ij}\bar{Q}_i Q_j + \kappa_{i\alpha}\kappa^\star_{j\beta}\bar{Q}_i \gamma_\mu Q_j \bar{d}_{R\alpha}\gamma^\mu d_{R\beta} + \mathrm{h.c.}
\end{equation}  
Taking universal masses and couplings for the dark pions, along with assuming $m_{\pi_d} > \Lambda_{QCD}$, the decay width of the dark pion to SM quarks is:
\begin{equation}\label{eqn:DPdecay}
\Gamma (\pi_d \rightarrow \bar{q}q) \approx \sum_q \frac{\kappa^4 N_c f^2_{\pi_d}m^2_q}{32\pi m_X^4}m_{\pi_d}, 
\end{equation}
with $N_c = 3$ being the number of SM colours, $m_q$ the mass of the SM quark, and $f_{\pi_d}$ being the dark pion decay constant. The sum in Eq.~\eqref{eqn:DPdecay} is over all kinematically accessible SM quarks: the largest kinematically accessible quark is the most important due to the factor of $m_q^2$ in the decay width, this arises due to spin-parity effects (similar to those witnessed in charged pion decay). This decay width results in a proper lifetime
\begin{equation}\label{eqn:Lifetime}
\begin{split}
c\tau \approx 80 \,\mathrm{mm} \times \frac{1}{\kappa^4}\left(\frac{2\,\mathrm{GeV}}{f_{\pi_d}}\right)^2 \\
\left(\frac{100\,\mathrm{MeV}}{m_q}\right)^2
\left(\frac{2\,\mathrm{GeV}}{m_{\pi_d}}\right)^2 \left(\frac{m_{X_d}}{1\,\mathrm{TeV}}\right)^4.
\end{split}
\end{equation}
This clearly leads to macroscopic decay lengths, permitting a parameter space that can be of interest to the LLP detectors discussed in Sec.~\ref{sec:detectors}. In particular if $\kappa \sim 0.1$, these dark pions will regularly travel 100's of meters required to get to even the furthest away detectors. As discussed below, the production rate is largely independent of $\kappa$. Dark pions that are kinematically forbidden from decaying to kaons result in a massive increase in lifetime --- crossing this threshold will increase the proper lifetime by a factor of $\sim 400$.

The main production channels at the LHC is pair production via QCD processes $q\bar{q} \rightarrow X_d X_d^\dagger$ and $gg \rightarrow X_d X_d^\dagger$. The $X_d$ then decay to a quark $q$ and a dark quark $Q_d$. The rate of this process at a given collider depends only on the mass of the $X_d$. 
Functionally, this production process is extremely similar to squark production (for a single flavour) in supersymmetry with the primary difference being a factor of $N_d$ due to the multiplicity from dark colour. With the HL-LHC expected to generate an integrated luminosity of $\sim 3000\, \mathrm{fb}^{-1}$ over its lifetime \cite{Schmidt:2016jra}, potentially tens of thousands of signal events can be produced. It should be noted that next-to-leading order corrections (including the process $pp \rightarrow j X_d X_d^\dagger$) can be significant, with the similar squark production indicating a K-factor of around 1.3 \cite{Beenakker:1996ch}. Since our focus is on producing a very broad overview of the potential reach of new LLP detectors, for simplicity we use tree-level cross sections for our analysis.

Another production process for dark quarks is a $t$-channel exchange of $X_d$ to give $q\bar{q}\rightarrow Q_d \bar{Q}_d$. This process depends on the mass of the $X_d$ and also the coupling $\kappa$ and is thus more model dependent. It may, however, be dominant if $m_X$ is large because it does not require two on-shell $X_d$ particles. The $t$-channel process also has different kinematics, with the dark quarks more likely to be produced in the forward direction, as opposed to pair-production which is more likely to be produced centrally. The main focus of our study will be pair production and detectors at relatively large angles, but in App.~\ref{sec:LLP} we explore the $t$-channel process focusing on more forward detectors FASER, FORMOSA, and MAPP.

The phenomenology of this model is essentially controlled by three variables:
\begin{equation}\label{eqn:KeyVariables}
m_{\pi_d}, \,\,\,\,\,\,\, \tau_{\pi_d}, \,\,\,\,\,\,\, m_{X}
\end{equation}
where the universal Yukawa coupling $\kappa$ has been replaced with the dark pion lifetime $\tau_{\pi_d}$. Of course, breaking the assumption of Yukawa universality  significantly increases the number of free parameters.  

The dark pion lifetime is of crucial importance in determining the type of signals these dark QCD models produce. Depending on the dark pion lifetime, there are three different final states that can be considered \cite{EJPlus}: 1) the dark pions are long-lived ($c\tau_{\pi_d} \gtrsim 1 \, \mathrm{m}$) and appear as missing energy in the primary LHC detectors, 2) the dark pions have intermediate lifetimes ($1 \, \mathrm{mm}\lesssim c\tau_{\pi_d} \lesssim 1 \, \mathrm{m}$) and produce Emerging Jets, and 3) the dark pions decay promptly ($c\tau_{\pi_d} \lesssim 1 \, \mathrm{mm}$), producing 4 jets. For this work, the first scenario is our primary interest as it is the chunk of parameter space most easily explored by LLP detectors. However, it should be noted that the intermediate parameter space with Emerging Jets can also be probed by these distant detectors, making them highly complementary to the LHC in this regime.

Missing energy, the CMS Emerging Jets search~\cite{CMS:2018bvr}, and prompt 4-jet searches have been used~\cite{EJPlus} to put exclusion bounds on the Eq.~\ref{eqn:KeyVariables} parameter space. The limits on the $X_d$ mass are a complicated function of dark pion mass and lifetime, but they are all on the order of 1 TeV. Therefore in this work we fix $m_{X} = 1000$ GeV. Because of the strong dynamics of the dark sector, the kinematics of the dark pions are a relatively weak function of $m_X$, and results for other values of $m_X$ can be approximated by rescaling the cross section.

\section{Long-Lived Particle Detectors}
\label{sec:detectors}

\begin{table*}
\begin{center}
\begin{tabular}{ |p{2cm}||p{3cm}|p{3cm}|p{2.5cm}|p{3cm}|p{2cm}|  }
 \hline
 \multicolumn{6}{|c|}{Overview of Examined Long-Lived Particle Detectors} \\
 \hline
 Experiment & Volume ($\mathrm{m}^3$) & Distance from IP (m) & Geometry & Luminosity ($\mathrm{fb}^{-1}$) & References\\
 \hline
 AL3X & $\sim 900$ & 4.25 & Cylinder & $\cO(100)$ & \cite{AL3X}\\
 \hline
  ANUBIS & $\sim 15*10^3$ & 0 & Cylinder & 3,000 & \cite{ANUBIS}\\
 \hline
  CODEX-b & $\sim 10^3$ & 25 & Box & 300 & \cite{CODEXb,Gligorov:2017nwh} \\
 \hline
  FASER & $\sim 15$ & 480 & Cylinder & 3,000 & \cite{FASER,FASER2,FASER3}\\
 \hline
  FORMOSA & $4$ & 500 & Box & 3,000 & \cite{FORMOSA}\\
 \hline
  MAPP & $\sim 150$ & 55 & Box & 30 & \cite{MAPP,MAPP2}\\
 \hline
  MATHUSLA & $\sim 2.5*10^5$ & $\sim 90$ & Box & 3,000 & \cite{MATHUSLA,MATHUSLA2,MATHUSLA3}\\
 \hline
  MilliQan & 3 & 33 & Box & 300 & \cite{MilliQan,MilliQan2}\\
 \hline
\end{tabular}
\caption{Overview of the long-lived particle detectors examined in this paper.}\label{tab:cases}
\end{center}
\end{table*}


Here we describe the landscape of different detectors that we explore. The critical values are summarized in Tab.~\ref{tab:cases}. 

\textbf{AL3X} \cite{AL3X}: \textit{A Laboratory for Long-Lived eXotics} experiment, AL3X, is a proposed LLP experiment that would be constructed within the L3 electromagnet and ALICE time projection chamber at the LHC, near interaction point 2. A cylindrical detector with $\sim 900\, \mathrm{m}^3$ volume, AL3X has quite a short baseline $\sim 5\, \mathrm{m}$ and heavy shielding from SM backgrounds. Given the hypothetical detector's considerable geometric acceptance, AL3X's reach extends beyond LLPs generated by  energy portals and also includes sensitivity to low scale vector, scalar or fermion mixing portals --- ultimately covering all possible renormalizable couplings of the SM to exotic sectors. 

\textbf{ANUBIS} \cite{ANUBIS}: \textit{AN Underground Belayed In-Shaft search experiment}, ANUBIS, is a proposed experiment that would utilize the 18 m in diameter, 56 m long PX14 installation shaft at ATLAS~\citep{Collaboration_2008}. This off-axis LLP experiment would instrument approximately $\sim 15,000\,\mathrm{m}^3$ with dedicated LLP detectors with sensitivity to lifetimes ranging from $0 \lesssim c\tau \lesssim 10^6 \; \text{m}$. A major advantage of ANUBIS is its potential to be synchronized and fully integrated with ATLAS, creating a continuous tracking volume from the interaction point to the top of the shaft 80 m away. With four tracking stations placed 18.5 m apart, the main detection volume would be divided into 3 different regions. Two $1 \,\mathrm{m} \times 1 \,\mathrm{m}$ tracking station prototypes (to be suspended at the top and bottom of PX14) have been proposed. 

\textbf{CODEX-b} \cite{CODEXb}: The \textit{COmpact Detector for EXotics at LHCb}, CODEX-b, experiment is a proposed detector that would search for transverse LLPs. The detector would be located in the DELPHI/UXA cavern next to LHCb's interaction point (IP 8). With dimensions of $10 \,\mathrm{m} \times 10 \,\mathrm{m} \times 10 \,\mathrm{m}$, CODEX-b would be a shielded, transverse, ``background-free" detector specializing in looking for light LLPs ($m_{LLP} \lesssim 10-100\, \mathrm{GeV}$) with long lifetimes ($1\lesssim  c\tau/\mathrm{m} \lesssim 10^7$) and high $\sqrt{\hat{s}}$ production channels. Additionally, the new experiment could be integrated with LHCb and be used to tag events of interest within LHCb. A prototype detector, CODEX-$\beta$~\cite{CodexbSnowmass} is proposed to be installed during LHC run 3. 

\textbf{FASER} \cite{FASER2}: FASER, the \textit{ForwArd Search ExepRiment} is an experiment (expected to begin taking data in 2022) located along the ATLAS beam collision axis, 480 m from the interaction point in service tunnel TI12. Sensitive to decays within a cylindrical volume with a radius of R = 10 cm and length L = 1.5 m, FASER can be used to constrain all types of dark photons, dark Higgs bosons, heavy neutral leptons, axion-like particles, and more. Broadly speaking, two types of target signals are being searched for: two oppositely charged tracks or two photons $\sim$ TeV energies that emanate from a common vertex within the detector and point back towards the interaction point. An additional neutrino detector FASER$\nu$~\cite{FASER:2020gpr} has been added, giving the experiment the ability to detect neutrinos. Providing that FASER is a success, a similar, larger detector (R = 1 m, L = 5 m), FASER 2~\cite{FASER2}, will be installed during the Long Shutdown 3 and collect data during the HL-LHC era.  

\textbf{FORMOSA} \cite{FORMOSA}: The proposed \textit{FORward MicrOcharged SeArch} experiment is a far-forward detector that would be located $\sim 500$ m downstream from the ATLAS interaction point in either cavern UJ12 or service tunnel TI12. FORMOSA's main objective is to cover a significant portion of the millicharged strongly interacting dark matter window, but also features the ability study additional BSM scenarios. The proposal suggests the development of an initial detector (FORMOSA-I) with dimensions $0.2 \,\mathrm{m} \times 0.2 \,\mathrm{m} \times 4 \,\mathrm{m}$ which would be followed up by a larger version (FORMOSA-II) with dimensions $1 \,\mathrm{m} \times 1 \,\mathrm{m} \times 4 \,\mathrm{m}$. 

\textbf{MAPP} \cite{MAPP}: A proposed subdetector of the MoEDAL experiment, the \textit{MoEDAL Apparatus for detecting Penetrating Particles} has two primary purposes: to search for millicharged particles and to observe decaying LLPs from renormalizable portal interactions. With dimensions of $5 \,\mathrm{m} \times 10 \,\mathrm{m} \times 3 \,\mathrm{m}$, the first-stage detector MAPP-1 would be movable within the UGC1 gallery; it could be deployed in configurations ranging from $5^\circ$ off the beamline at a distance of 55 m from IP8, to $25^\circ$ off at a distance of 25 m. Current plans will see MAPP-1 installed and ready for data taking during LHC Run 3. A further upgrade to a larger version of the detector, MAPP-2, is envisioned for the HL-LHC in 2026.

\textbf{MATHUSLA} \cite{MATHUSLA3}: The \textit{Massive Timing Hodoscope for Ultra-Stable neutraL pArticles} is the largest of the proposed LLP detectors, with a decay volume coming in at $100 \,\mathrm{m} \times 100 \,\mathrm{m} \times 25 \,\mathrm{m}$ at $\sim 90$ m from the CMS interaction point. It should be noted that the original letter of intent the dimensions were 200 m $\times$ 200 m --- the newer version is closer to the interaction point and features extremely similar LLP sensitivity for only a quarter of the cost. Due to its location at the surface, there are backgrounds from both cosmic rays along with LHC produced muons. However, measurements done by the MATHUSLA test stand in 2018, coupled with detailed simulations studies confirm earlier estimates that downward traveling cosmic rays, muons from the LHC and atmospheric neutrinos can be vetoed and are unlikely to constitute a background to LLP searches at MATHUSLA. The collaboration aims to produce a full technical design report in 2022 and have MATHUSLA operational when the HL-LHC comes online in $\sim2026$.

\textbf{MilliQan} \cite{MilliQan}: A dedicated millicharged particle detector, MilliQan is a proposed experiment that would be placed in the PX56 drainage gallery above CMS UCX and would detect particles produced at LHC point 5. The detector concept is a $1 \,\mathrm{m} \times 1 \,\mathrm{m} \times 3 \,\mathrm{m}$ plastic scintillator array oriented such that the long axis points towards the CMS IP 33 m away. A 1/100th scale version was installed by the collaboration in 2018: this feasibility prototype for MilliQan successfully ran under $37 \, \mathrm{fb}^{-1}$, collecting information for full detector installation, event triggering, time and charge calibration.

\begin{figure}[h]
\centering
\includegraphics[width=0.5\textwidth]{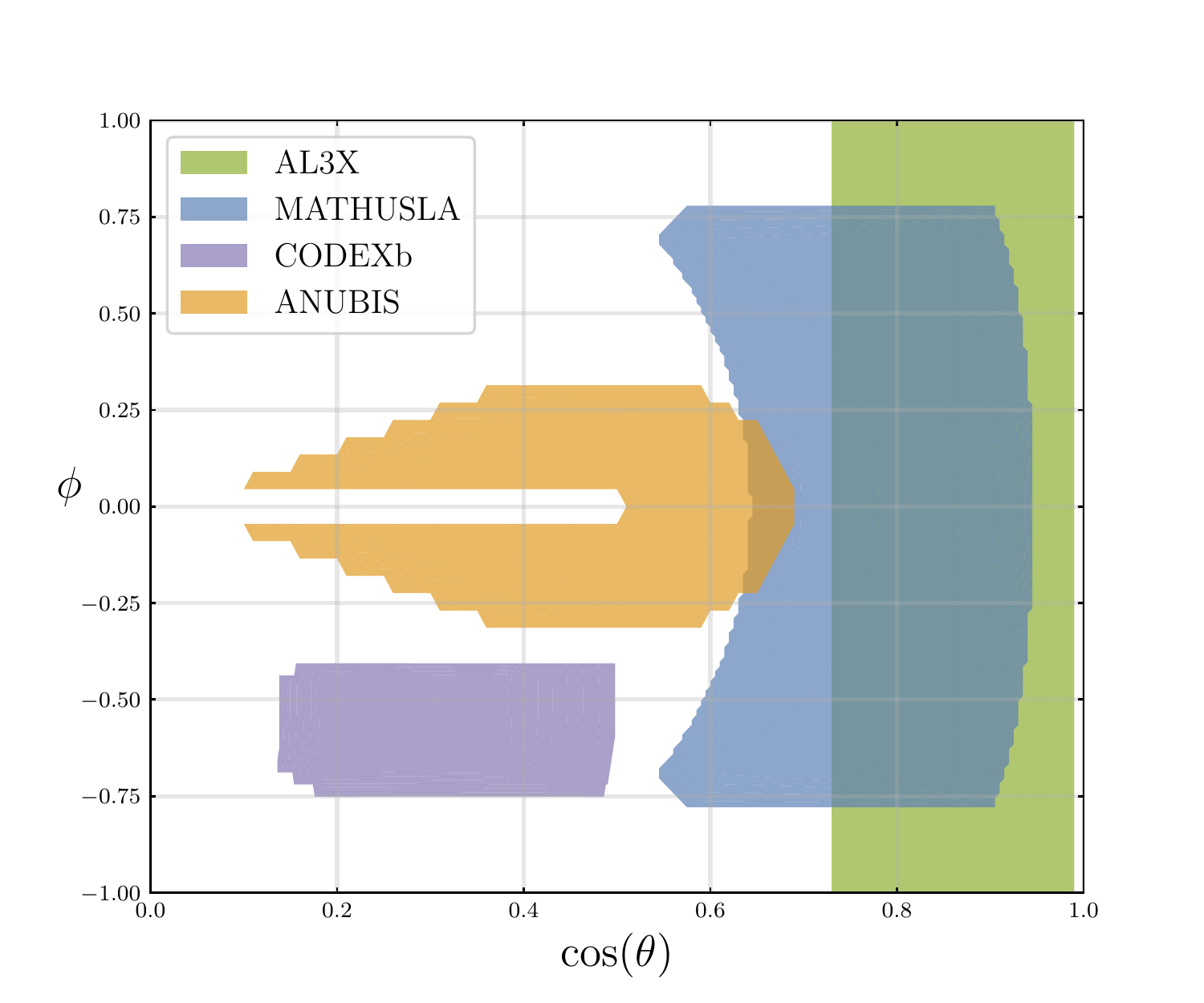}
\caption{Detector Coverage of the four most sensitive LLP experiments in angular space. Each area was estimated using the detector simulation of Section~\ref{sec:events}. To achieve a clearer picture, CODEX-b was shifted in $\phi$  by $-\pi/2$ which is justified by the physics being $\phi$-independent. AL3X covers the entire $\phi$ range ($0, 2\pi$). }
\label{Fig:cov}
\end{figure}

As we will show, the detectors with the best discovery potential for Emerging Jets are AL3X, ANUBIS, CODEX-b, and MATHUSLA. We give a comparison of the geometric coverage in $\theta-\phi$ space of those four detectors in Fig.~\ref{Fig:cov}. 

\section{Event Generation and Simulation}
\label{sec:events}

\begin{figure*}
\centering
\begin{minipage}[c]{\textwidth}
\includegraphics[width=0.495\textwidth]{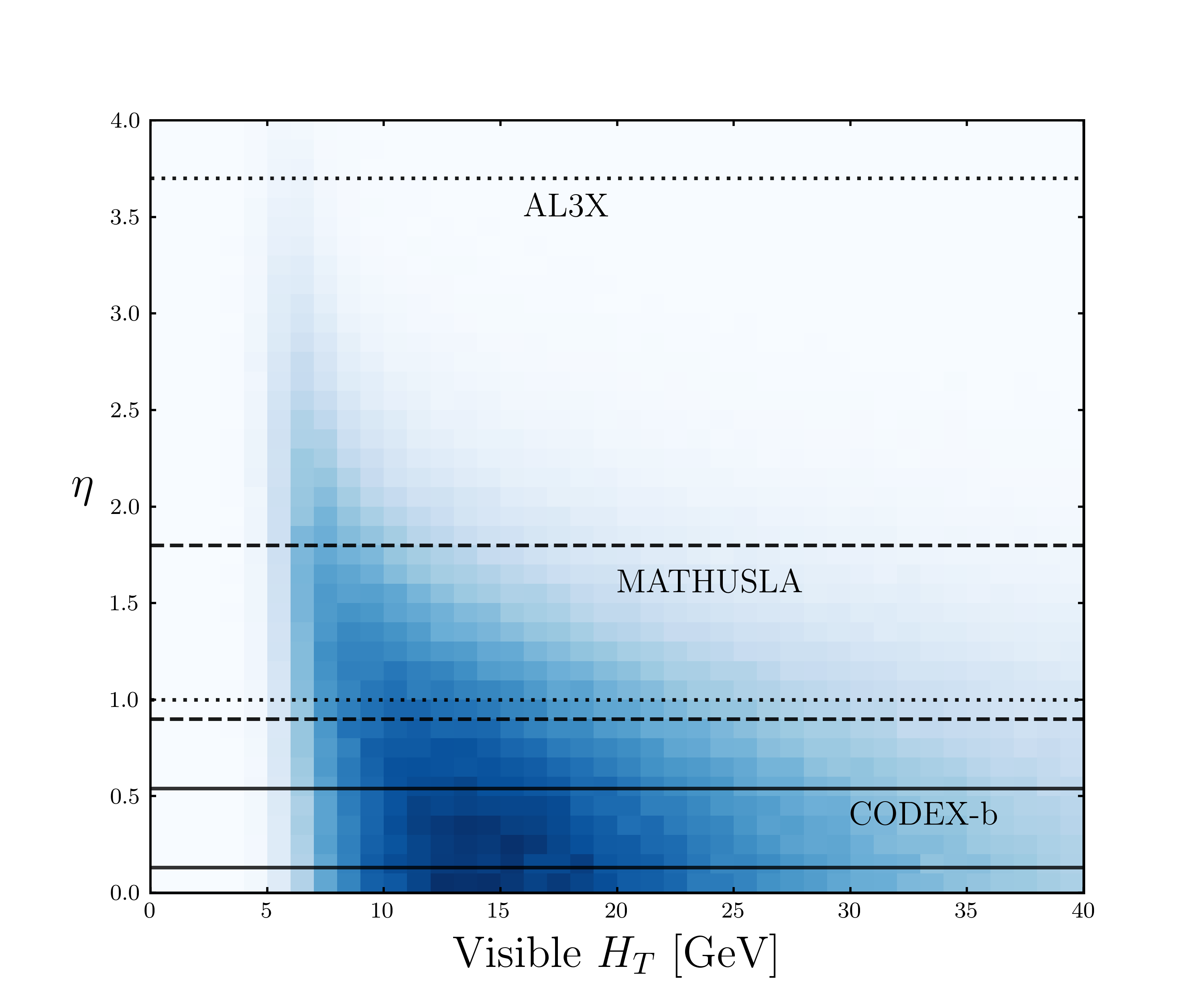}
\hfill
\includegraphics[width=0.495\textwidth]{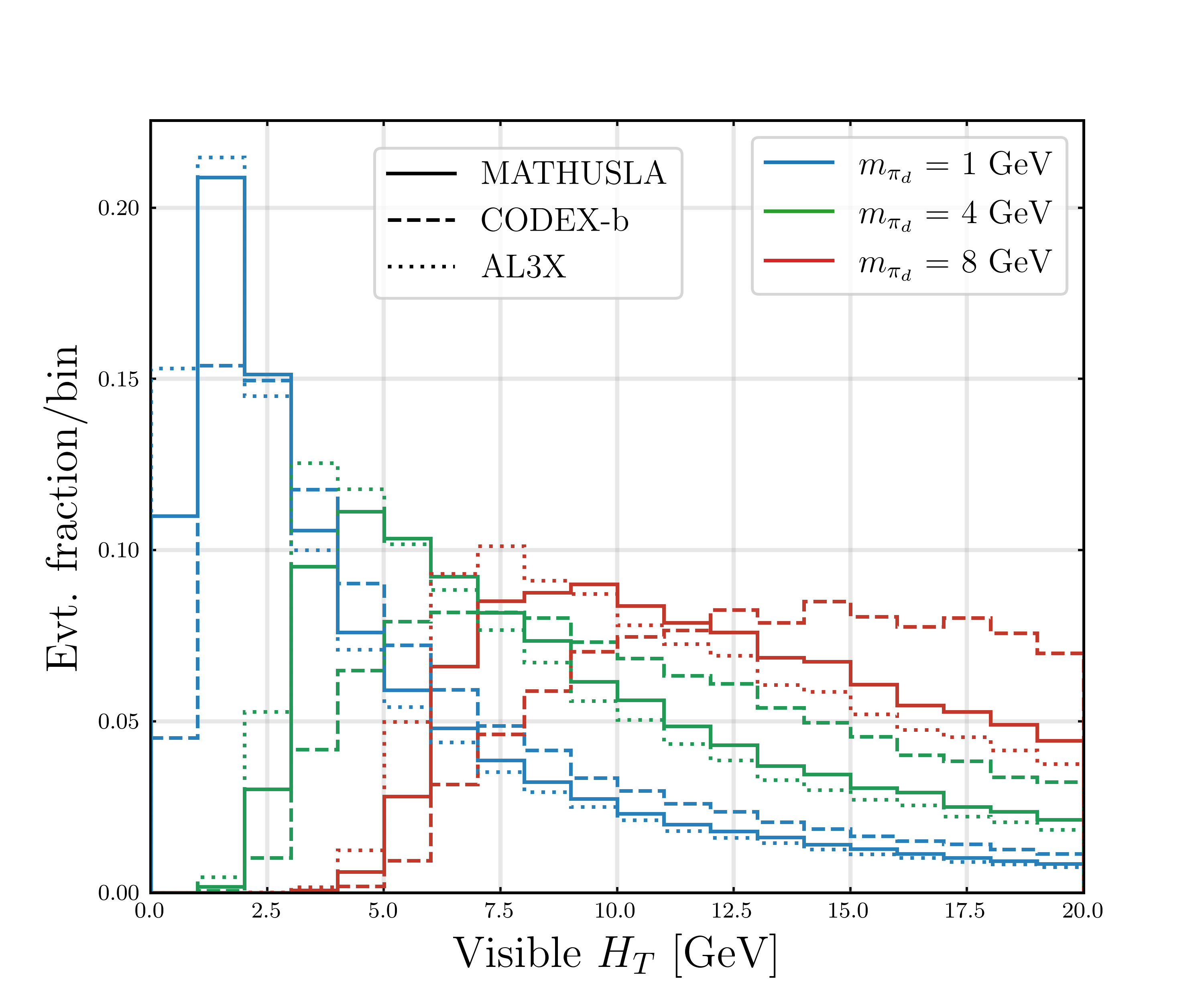}
\end{minipage}
\hfill
\caption{The scalar $p_{T}$ sum of visible tracks ($H_{T}$) resulting from decayed dark pions. Left: Transverse LLP angular acceptances for visible $H_T$ using the $m_{\pi_d} $ = 8 GeV. Right: visible $H_{T}$ distribution for three different dark pion benchmarks: $m_{\pi_d}=1$ GeV (softest spectrum in blue), $m_{\pi_d}=4$ GeV (intermediate spectrum in green), $m_{\pi_d}=8$ GeV (hardest spectrum in red). Different line styles correspond to different $\eta$ region associated with the detector's acceptance, solid is MATHUSLA, dashed is CODEX-b, and dotted is AL3X.   }
\label{Fig:Ht}
\end{figure*}

Here we describe our simulation pipeline to generate Monte Carlo events for estimating the detector acceptances. The events were generated using \verb!Pythia8!~\cite{Sjostrand:2014zea} under the LHC conditions with a center of mass energy of $\sqrt{s} = 13$ TeV. To model the pair production of the bifundamental $X_{d}$, \verb!Pythia8!'s hidden valley (HV) module~\cite{Carloni_2010,Carloni_2011} was used. Specifically, the production processes modelled are $g\bar{g} \rightarrow D_{v}\bar{D}_v$ and $q\bar{q} \rightarrow D_{v}\bar{D}_v$, where $D_v$ is given the same quantum numbers as the bifundamental with mass $m_X = 1$ TeV.  The showering of both QCD and dark QCD are initiated within \verb!Pythia8!, producing heavy dark states that either become stable or decay into dark pions $\pi_{d}$. The dark pions will travel macroscopic lengths, eventually decaying back into the SM through the bifundamental portal.

The spectrum of the HV is taken to be similar to the scheme in \cite{EJ, EJTrig, Mies:2020mzw} where
\begin{equation}\label{eq:dark_spectrum}
m_{Q_{d}} = \Lambda_{d} = 2m_{\pi_d} = \frac{1}{2}m_{\rho_d}.
\end{equation}
It should be noted that the quark mass is the constituent quark mass and not that in the Lagrangian in Eq.~(\ref{eqn:Lagrangian}). The dark vector mesons $\rho_d$ are made to decay promptly to dark pions --- which are, in turn, set to predominantly decay into strange quark pairs, which are further hadronized through \verb!Pythia8!. 

For each event, the number of dark pions is recorded, accompanied with their boost, azimuthal angle, and polar angle $(b,\phi,\theta)$. These kinematical variables are used to reconstruct lifetimes, decay lengths, and angular acceptance. Following the procedure in \cite{Dercks:2018eua, Hirsch:2020klk}, the total number of decayed dark pions can be estimated using

\begin{equation}
 N_{\pi_{d}}^{dec} = \epsilon \cdot N_{\pi_{d}}^{prod} \cdot \big \langle P(\pi_d \text{ in d.r}) \big \rangle \cdot \text{BR}(\pi_{d} \rightarrow \text{signal}),
\end{equation}
where $\text{BR}(\pi_{d} \rightarrow \text{signal})$ is the branching ratio of the decay into visible channels, $N_{\pi_{d}}^{prod}$ is the total number of dark pions produced, $\epsilon$ is the efficiency of reconstructing the signal tracks, and $\big \langle P(\pi_d \text{ in d.r}) \big \rangle$ is the detector efficiency of having a decay within the detector region. The number of dark pions produced was calculated using the expected total integrated luminosity for each detector from Tab.~\ref{tab:cases} using 

\begin{equation}
N_{\pi_{d}}^{prod} = \mathcal{L} \cdot \sigma(pp \rightarrow X_{d}).
\end{equation}
Here, $\sigma(pp \rightarrow X_{d})$ is the cross section of producing the scalar bi-fundamental. The cross section is related to squark production as they share the same quantum numbers. Using 13 TeV calculations from~\cite{Beenakker:2016lwe}, the cross section of $\sigma(pp \rightarrow X_{d}) = N_{d}\times6.2$ fb was obtained for $m_X = 1$ TeV, and we take $N_d = 3$. The bi-fundamental decays purely into dark pions and dark vectors, which, as alluded to earlier, promptly decay into dark pions. Thus, the branching ratio into dark pions is taken to be $100\%$.

The detector acceptance was estimated using Monte Carlo simulations for each of the proposed LLP detectors. Simple mock detector geometries were constructed using the specifications in Section~\ref{sec:detectors}, creating a fiducial detector acceptance region\footnote{The 
code is publicly available at \href{https://github.com/DLinthorne/LLP-Experiments}{on github} 
and described in App.~\ref{sec:DS}.}. The walls of each detector are taken to be infinitesimally thin. Therefore the energy loss due to material interactions is neglected to simplify the models. The expected detector efficiency is
\begin{equation}\label{eq:monte}
\big \langle P(\pi_d \text{ in d.r}) \big \rangle = \frac{1}{N_{\pi_d}^{MC}} \sum_{i = 0}^{N_{\pi_{d}}^{MC}} P((\pi_d)_i \text{ in d.r}),
\end{equation}
where $N_{\pi_d}^{MC}$ is the total number of dark pions simulated. The r.h.s of Eq.~(\ref{eq:monte}) sums the probability of each the $i^{th}$ dark pion decaying within the detector region. The probability is simply
\begin{equation}
P((\pi_d)_i \text{ in d.r}) = e^{-L_1/\lambda_i}\cdot(1 - e^{L_2/\lambda_i}),
\end{equation}
where $\lambda_{i}$ is the decay length of the $i^{th}$ dark pion computed from its boost $\lambda = \gamma \beta c \tau$. $L_2$ ($L_1$) is the distanced travelled within (to reach) the detector region. In ~\cite{Dercks:2018eua, Hirsch:2020klk} $L_1$ and $L_2$ were analytically approximated for each LLP detector. Instead, here $L_{1}$ and $L_{2}$ are calculated using the kinematical information of each dark pion with respect to the constructed detector geometry. 

This Monte Carlo detector simulation creates a more realistic model of detector acceptance. The proposed positions and shapes of these detectors each take up unique and non-trivial areas within the ($\phi,\theta$) angular space. This can be seen in Fig.~\ref{Fig:cov}, which is the estimated geometric range of each detector generated by the detector simulation. FASER, FORMOSA, and MAPP are highly forward and thus not represented in Fig.~\ref{Fig:cov}.

\section{Results}
\label{sec:results}

\begin{figure*}
\centering
\begin{minipage}[c]{\textwidth}
\includegraphics[width=0.495\textwidth]{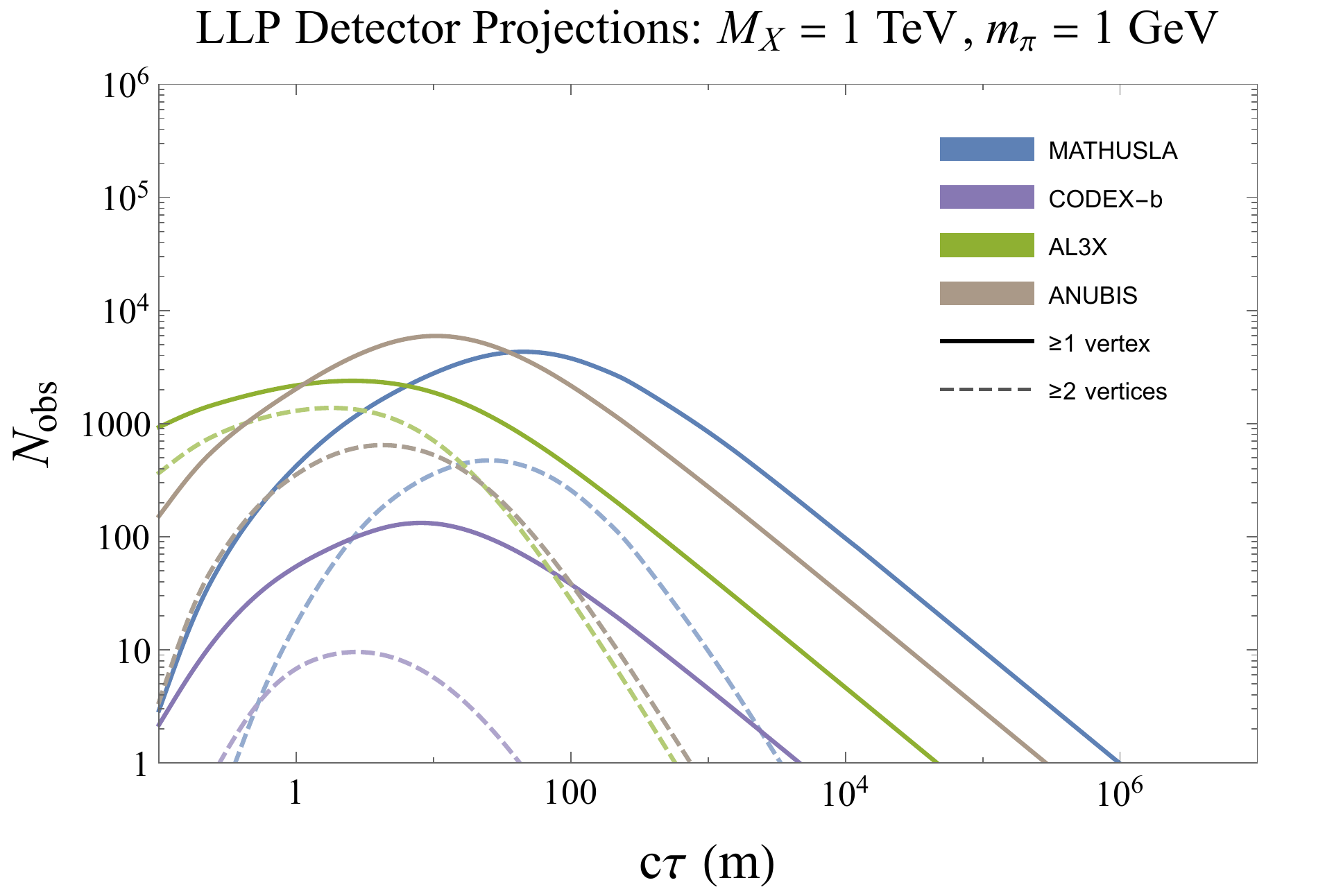}
\hfill
\includegraphics[width=0.495\textwidth]{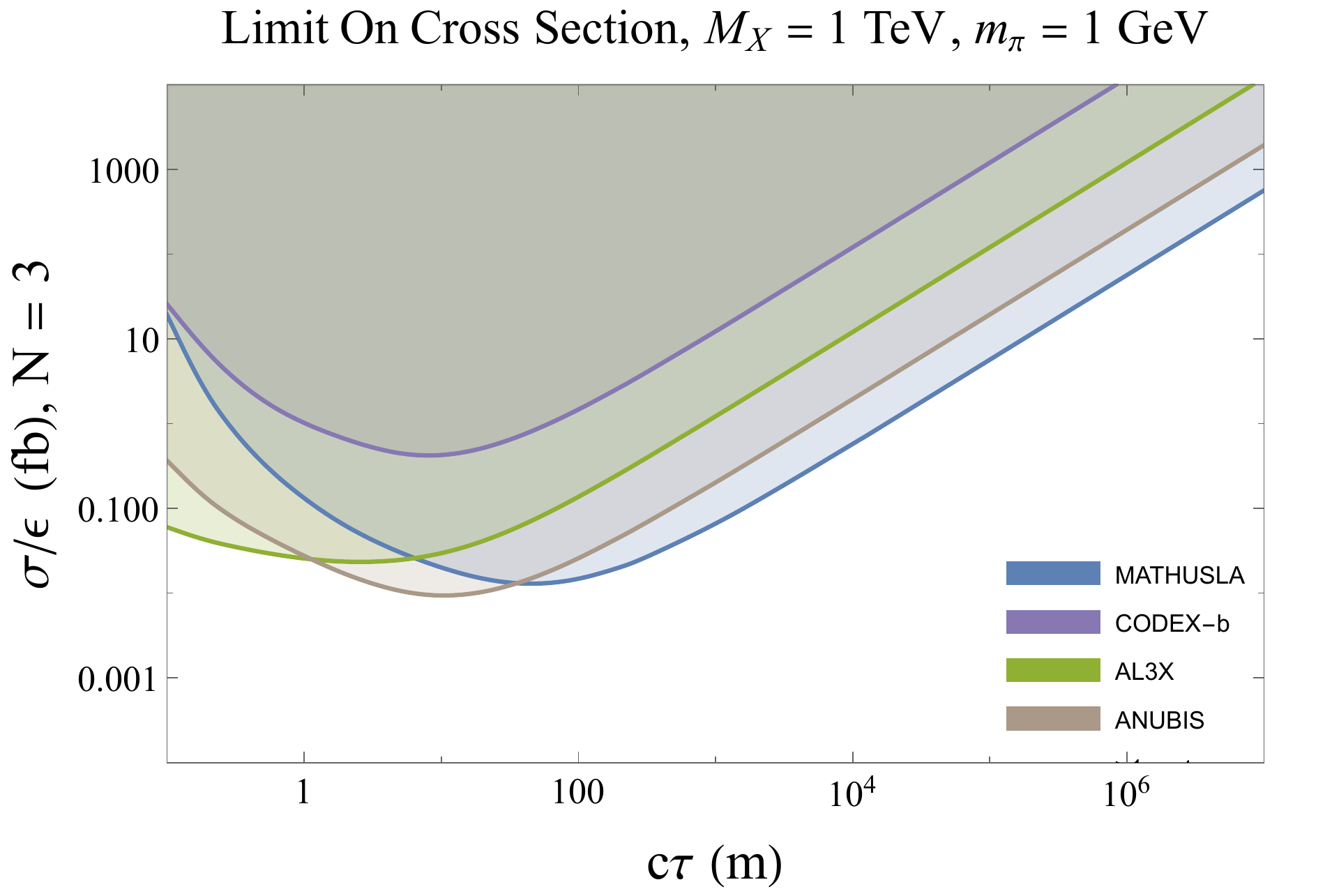}
\end{minipage}
\hfill
\caption{Left: expected number of observed dark pion decays observed at the LLP experiments as a function of dark pion lifetime. Solid (dashed) lines correspond to observing one (two) or more decays within the detector region. Right: the $95\%$ confidence upper bounds on the cross section per reconstruction efficiency $\epsilon_{DV}$, assuming a completely background-free experiment. Both figures are projections for $m_\pi$ = 1 GeV and $M_{X_d} = 1$ TeV. }
\label{Fig:lifetime}
\end{figure*}

\begin{figure}[b]
\centering

\includegraphics[width=0.5\textwidth]{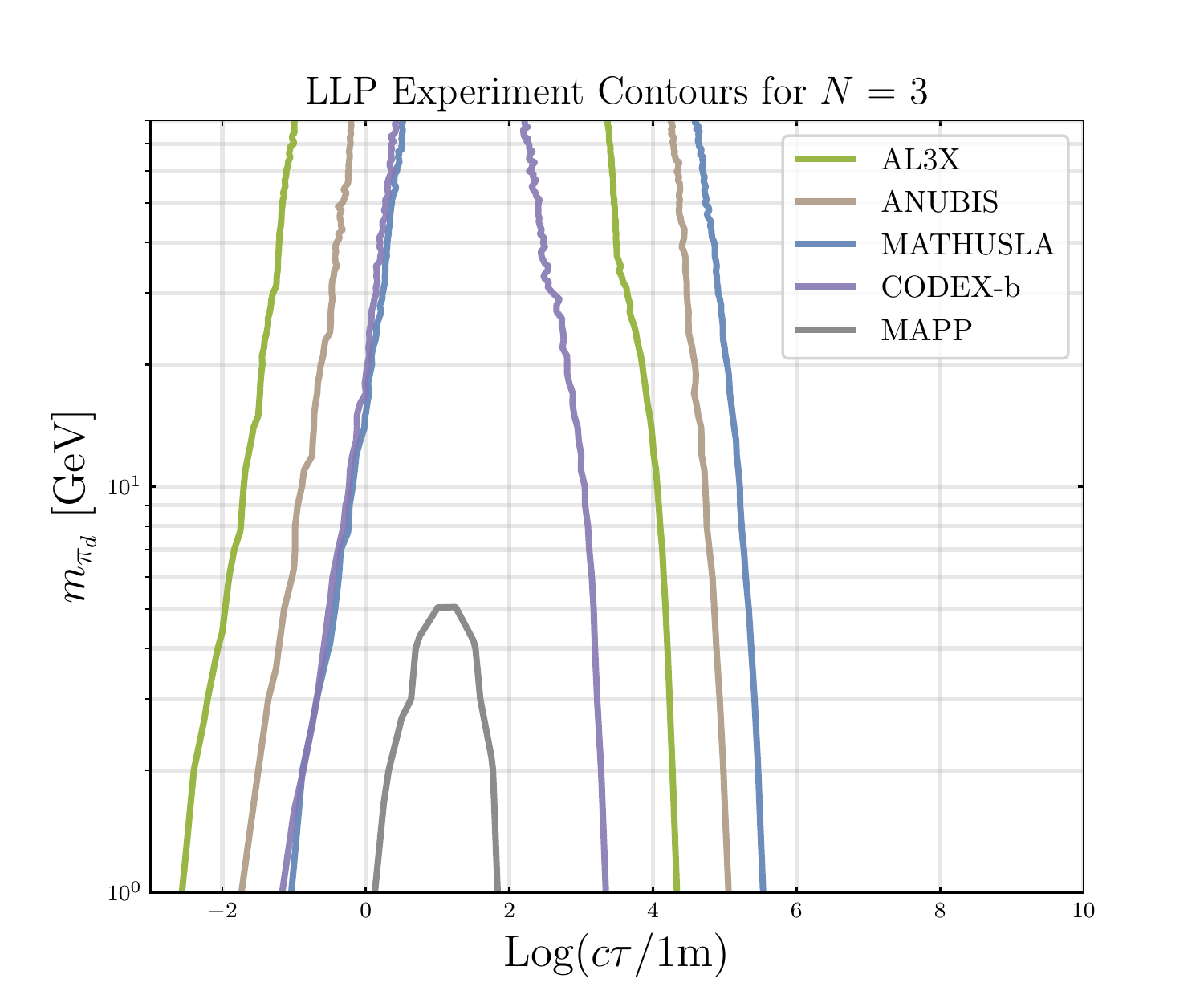}

\caption{Experimental sensitivity $95\%$ confidence interval, assuming a completely background-free experiment. These contours assume a fixed mediator mass of $M_{X_d} = 1$ TeV.}
\label{Fig:contour}
\end{figure}

Each LLP experiment quotes an idealized background-free experiment, which may be realized in some detectors more easily than others. To make the assumption of a background-free experiment more accurate, cuts are imposed on reconstructed visible tracks within the detector region, lowering the overall reconstruction efficiency of a displaced vertex $\epsilon$. CODEX-b will employ a $p_{T} > 400$ MeV cut on each track in an effort to reduce secondaries produced through material interactions. Emerging Jets have characteristically high multiplicities which may lower the average momentum of visible tracks. AL3X will also impose a cut of $H^{\text{vis}}_{T} = \sum|p_{T,i}| >$ 3 GeV, where the sum is over all visible tracks emanating from a single displaced vertex. Fig.~\ref{Fig:Ht} presents how $H^{\text{vis}}_{T}$ is distributed over the angular regions of the detectors given a fixed mediator mass $M_{X_d} = 1$ TeV.  Lower pseudorapidity will impart more visible energy in the event, which CODEX-b benefits the most from. In the right panel of Fig.~\ref{Fig:Ht} the distribution of $H^{\text{vis}}_{T}$ at MAHTHUSLA, CODEX-b, and AL3X are shown for three benchmark models. The total visible energy lowers for lower dark pion masses. This is a consequence of smaller pion masses producing higher multiplicities, therefore producing a lower number of visible tracks per pion. Explicitly, this means that smaller quark masses worsen the overall reconstruction efficiency $\epsilon$ at AL3X.

Using the event generation methods and detector simulation outlined in Section~\ref{sec:events}, the number of observed events $N_{\text{obs}}$ was estimated for the dark pion lifetime range of ($10^{-2}$ m - $10^{7}$ m) at each detector. In the left panel of Fig.~\ref{Fig:lifetime}, $N_{\text{obs}}$ is broken into two categories; the number of decays observed with one or more displaced vertices and the number of decays observed with two or more. MATHUSLA, ANUBIS, and AL3X are an order of magnitude more sensitive than CODEX-b. MATHUSLA and ANUBIS's high sensitivities are due to their higher integrated luminosities, as they will be using the full runs at the HL-LHC. Although AL3X and CODEX-b have similar integrated luminosities $\mathcal{O}(100 \,\text{fb}^{-1})$, AL3X covers a much larger angular acceptance, as seen in Fig.~\ref{Fig:cov}. These projections do not consider the reconstruction efficiency $\epsilon$, which enhances CODEX-b with respect to the other experiments. The detectors peak at different lifetimes due to their distance to the IP. 

The right panel of Fig.~\ref{Fig:lifetime} represents the $95\%$ confidence limits on the experimental sensitivities, assuming a completely background-free experiment. In the case that the background-free assumption is not achieved, the bounds represent the cross sections that correspond to measuring at least three signal events. 

To explore the dark sector parameter space further, we simulated a grid of of dark pion lifetime $c\tau$ and mass $m_{\pi_{d}}$, in steps of   1 GeV for $m_{\pi_{d}}$. A minimum dark pion mass value of 1 GeV is taken to be consistent with Eq.~(\ref{eqn:Lifetime}), which requires $m_{\pi_{d}} \geq \Lambda_{\text{QCD}}$. 10k events were simulated for each grid point, using the same methods highlighted in Section~\ref{sec:events}. In Fig.~\ref{Fig:contour} the estimated sensitivities are cast for each experiment. Again, these sensitivities represent a total of three signal events during the run, corresponding to a $95\%$ confidence interval of a background free-experiment. Similar to Fig.~\ref{Fig:lifetime} the experiments cover different lifetime regions depending on their distance to the IP. The lifetime regions are reduced at higher dark pion mass, due to the dark pion's boost lowering as its mass increases. 

\begin{figure*}
\centering
\includegraphics[width=1.0\textwidth]{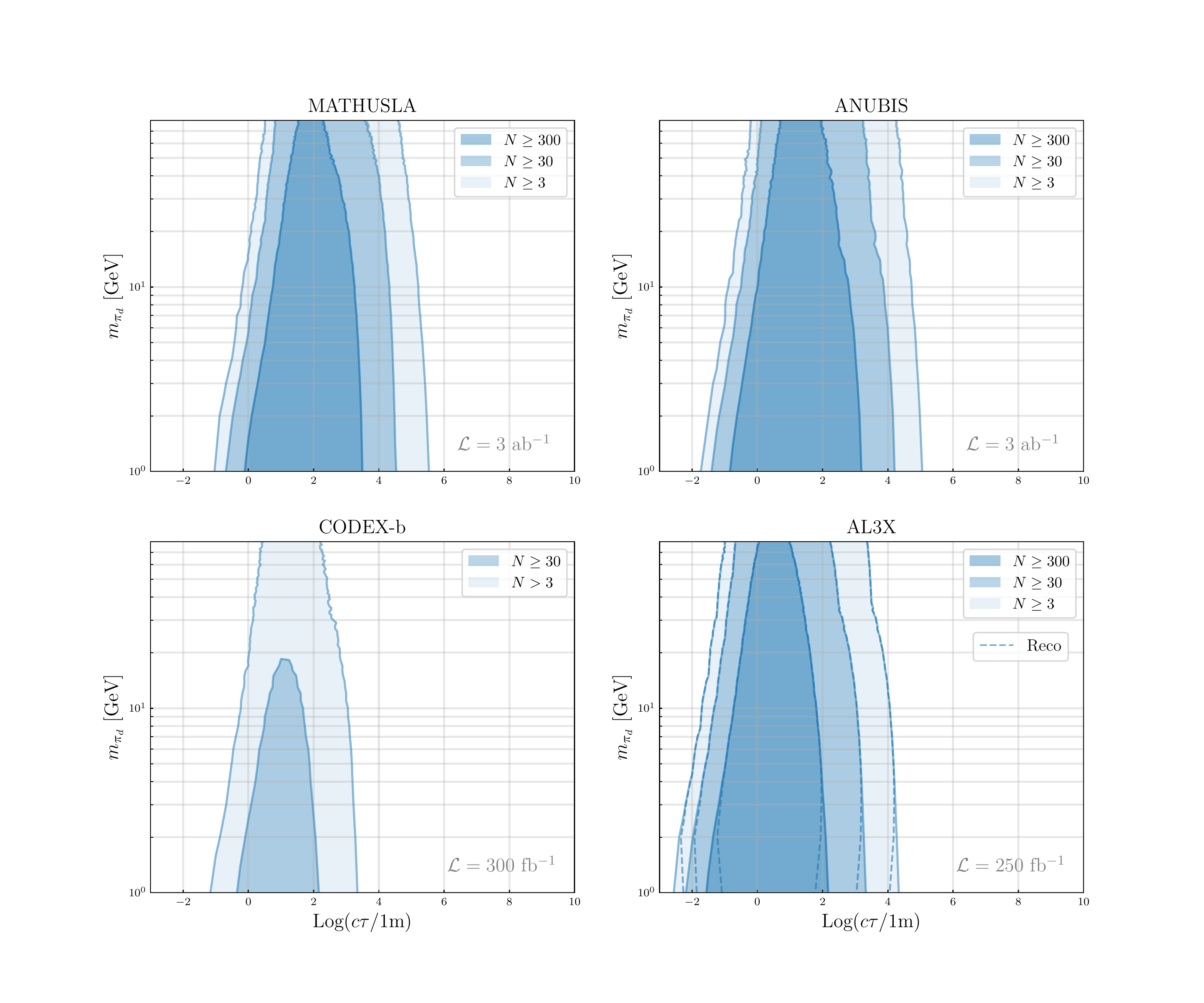}

\caption{Experimental contours of detecting $N >$ 3 (light blue), 30 (blue), 300 (dark blue) for the transverse experiments MATHUSLA, AL3X, CODEX-b, and ANUBIS. The dashed lines in the bottom right plot include the reconstruction (reco) efficiency $\epsilon \neq 1$. These contours assume a fixed mediator mass of $M_{X} = 1$ TeV.}
\label{Fig:contour_all}
\end{figure*}

In Fig.~\ref{Fig:contour_all} the four transverse experimental sensitivities are separated showing reaches for more than just $N = 3$, but also for $N =$ 30, 300, allowing relaxations on the background assumptions. The dashed lines in the bottom right panel include the proposed reconstruction cut for AL3X of $H^{\text{vis}}_{T} >$ 3 GeV. This means that the contours are of $N_{\text{obs}} \cdot \epsilon$, but we see this is only important at small dark pion mass. Additionally, it should be noted that the remaining solid lines in Fig.~\ref{Fig:contour} and \ref{Fig:contour_all} assume $\epsilon = 1$.

Across the board, AL3X, ANUBIS, CODEX-b, and MATHUSLA have comparable sensitivities, whereas the other four experiments we studied have significantly worse sensitivities. MilliQan is not sensitive largely because of smaller angular coverage, and FASER, FORMOSA and MAPP are very forward so they have very small efficiency for the pair production process compared to the transverse experiments. This snapshot, however, fails to tell the entire story and it should be noted that the forward experiments become more sensitive using a $t$-channel exchange process --- this is explored in Appendix~\ref{sec:LLP}.

\section{Conclusions}
\label{sec:conclusion}


Emerging Jet phenomenology is well motivated, and an experimental program to search for Emerging Jets is in its infancy. The generic predictions of these models is long-lived states, but how long-lived these states are can vary by orders of magnitude. If the lifetimes are relatively short, then the current detectors at the LHC are likely best suited for discovery, but for longer lifetimes, the suite of proposed experiments may be superior. 

The work presented here begins the process of quantitatively exploring the discovery reach of proposed experiments at the LHC, by showing the geometric acceptance of these detectors at the LHC along with the sensitivity of these detectors to Emerging Jet models. AL3X, ANUBIS, MATHUSLA, CODEX-b, and MAPP are the only proposed LLP detectors that can begin to probe the parameter space (although FASER, FORMOSA, and MAPP do have some sensitivity to $t$-channel production processes). Further, even among this group AL3X, ANUBIS, and MATHUSLA really shine --- each of these experiments has a large region of geometric acceptance that is not covered by any other detector. As a result, each of these detectors are the strongest at probing some part of the Emerging Jet parameter space: AL3X does best at shorter decay lengths $ < 1$ m, ANUBIS at middle distances $\sim 1-100$ m, and  MATHUSLA for anything longer-lived.

Perhaps the most important result of this analysis is that Emerging Jet models will produce enough signal events within LLP detectors such that even if backgrounds are high enough to require 30 (or even 300) signal events for detection, LLP experiments (specifically MATHUSLA, ANUBIS, AL3X, and CODEX-b) will still be able to probe the dark QCD parameter space. Even under these extremely conservative requirements, the various experiments will cover lifetimes spanning more than 5 orders of magnitude. 

\begin{figure*}
\centering
\begin{minipage}[c]{\textwidth}
\includegraphics[width=0.495\textwidth]{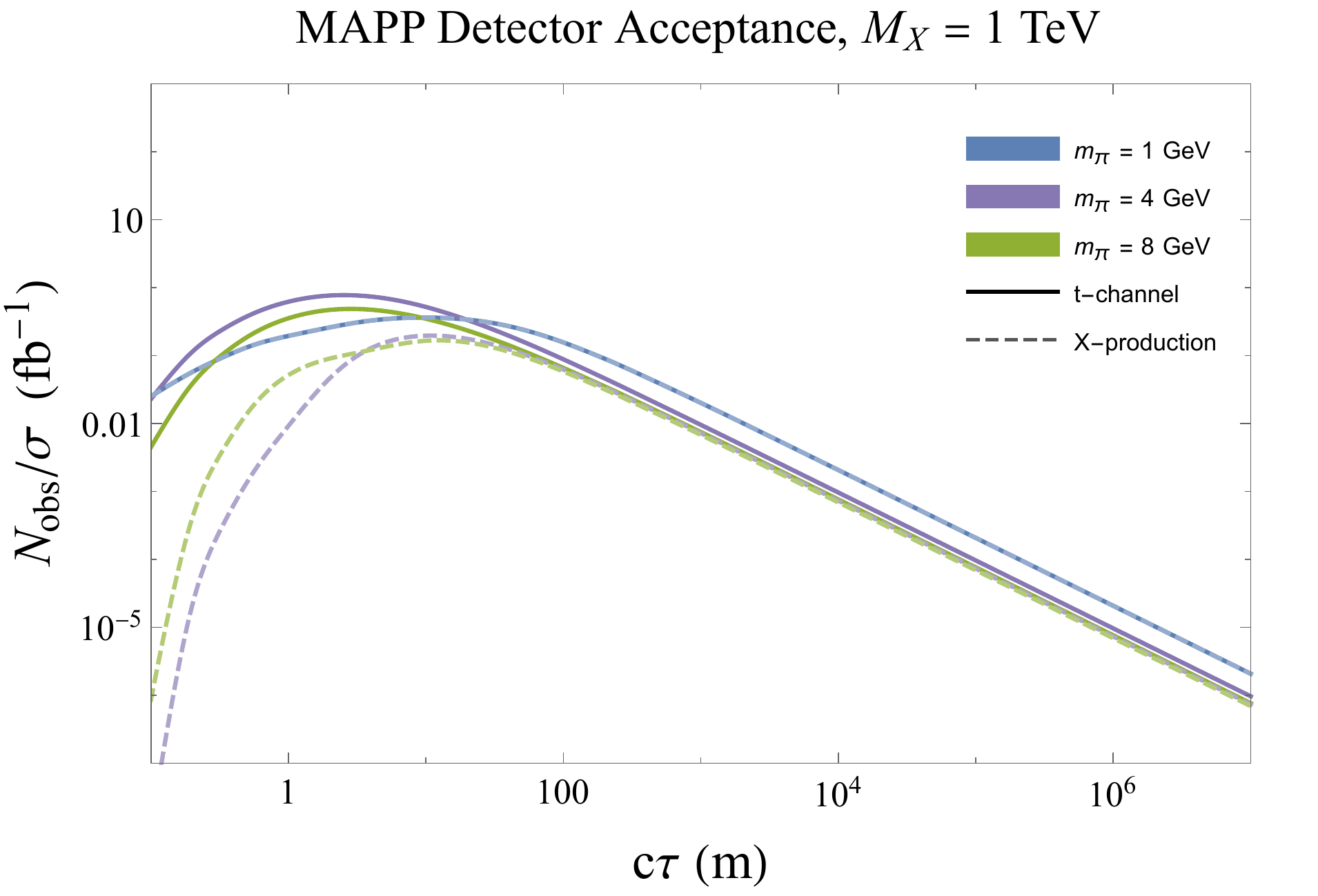}
\hfill
\includegraphics[width=0.495\textwidth]{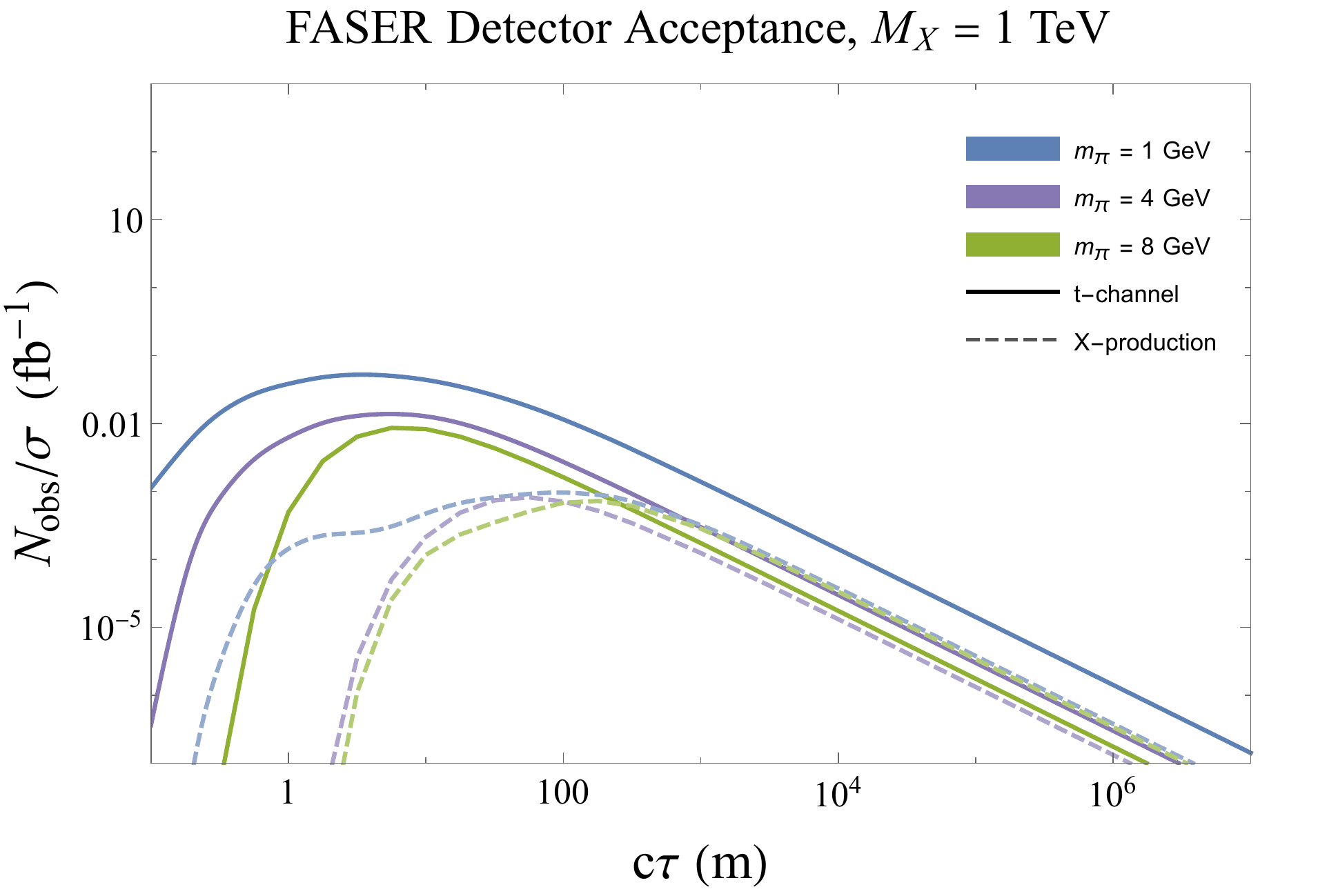}
\end{minipage}
\hfill
\caption{Expected observed dark pions per cross section for MAPP (left) and FASER (right) as a function of dark pion lifetime. Coloured lines distinguish between different pion mass benchmarks. Dashed lines represent the $X_d$-production explained in Section~\ref{sec:events}. Solid lines represent the $t$-channel exchange of $X_d$. }
\label{Fig:forward}
\end{figure*}

All-in-all, this work provides yet another reason to support the development of LLP experiments --- and several of them at that. These detectors cover intriguing dark QCD parameter space and, due to the huge range of possible lifetimes and geometric limitations of each individual detector, rely on each other to cover a much wider range of possibilities than would be accessible individually. Further, the regions where these detectors do overlap also serve an important role: the experiments can be used to cross-validate each other in these regions and help avoid potential systematic errors. 

\appendix

\section{$t$-channel projections}
\label{sec:LLP}

In Section~\ref{sec:events}, events were generated assuming pair production of the bifundamental mediator $X$. This proved to be the production channel of choice for the transverse experiments at small $\eta$, where the forward experiments at large $\eta$ had little to no sensitivity in the parameter spaces considered. Here we show how the $t$-channel exchange of $X$ can enhance the projected sensitivities of both MAPP and FASER. 

Currently, \verb!Pythia8! does not have the $t$-channel exchange of $X$ built in, therefore events were generated from the ground up using a modified $t$-channel model~\cite{Cohen_2015, Cohen_2017} implemented using the \verb!FeynRules!~\cite{Alloul_2014} package. The model is outputted as a UFO~\cite{Degrande_2012} file which allows generation of hard processes with \verb!Madgraph5_aMC@NLO!~\citep{Alwall:2014hca}, and we use LHC conditions with a centre of mass energy of 13 TeV. The dark quark PDG codes were implemented in \verb!Madgraph5_aMC@NLO! and their masses hard coded. As in Section~\ref{sec:events}, we fix $m_X = 1$ TeV and simulate a grid of pion masses from [1 GeV, 80 GeV] and liftetimes from [$10^{-2}$ m, $10^{-7}$ m], with 100k events at each grid point. This output is interfaced to the Hidden Valley~\cite{Carloni_2010,Carloni_2011} module of \verb!Pythia8!~\citep{Sjostrand:2014zea} discussed in Section~\ref{sec:events}. The resulting event's kinematical variables of interest $(\theta, \phi, b)$ were used to estimate the detector acceptance and subsequently the number of projected observations $N_{obs}$. 

\begin{figure}[h]
\centering

\includegraphics[width=0.495\textwidth]{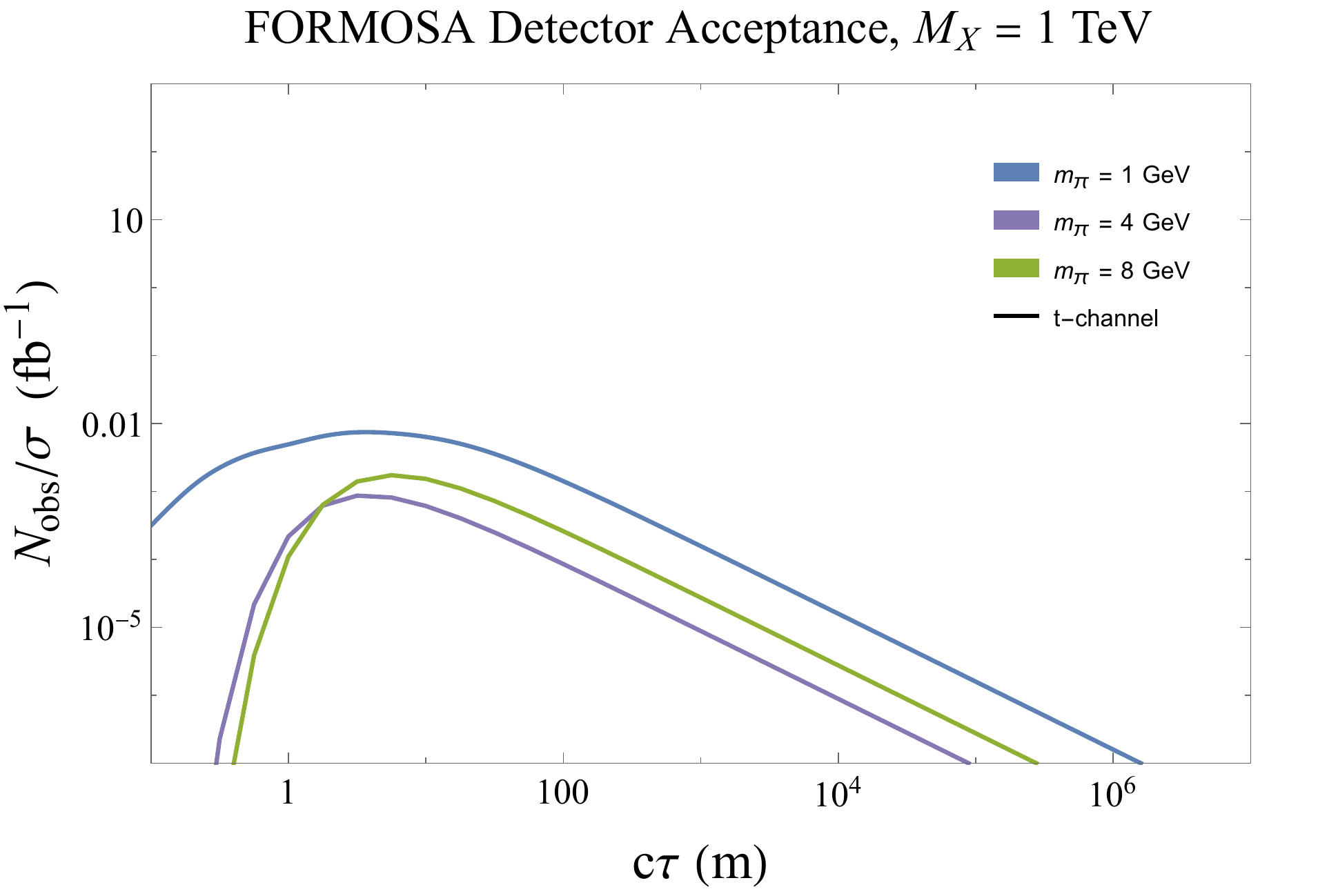}

\caption{Expected observed dark pions per cross section for FORMOSA as a function of dark pion lifetime. Coloured lines distinguish between different pion mass benchmarks. Solid lines represent the $t$-channel exchange of $X_d$. }
\label{Fig:formosa}
\end{figure}

Unlike pair-production, the $t$-channel cross section is dependent on the unknown coupling between the dark and visible sector. Therefore, in order to compare the two processes, $N_{obs}/\sigma = \mathcal{L} \cdot \big \langle P(\pi_d \text{ in d.r}) \big \rangle $ was calculated and plotted for both MAPP and FASER in Fig.~\ref{Fig:forward}. Both experiments have the $t$-channel process enhanced relative to pair-production, and FASER receives a greater enhancement of $\mathcal{O}(10^{3})$ due to it being more forward. Sensitivities are also projected for FORMOSA an equally forward detector. Only the $t$-channel process was estimated for FORMOSA in Fig.~\ref{Fig:formosa}, which has similar sensitivity to FASER.

\section{Detector simulation}
\label{sec:DS}

The detector simulation used in Section~\ref{sec:events} consists of simple geometric models of each of the detectors consider. We have made the models, source code, notebooks, and resources public so that they may be used for other studies and comparisons of the LLP detectors, and all the code can be found at  \href{http://github.com/DLinthorne/LLP-Experiments}{github.com/DLinthorne/LLP-Experiments}. Alongside the detector simulations, the repository hosts all model files and Monte Carlo simulations used to produce dark showering events.

The repository contains Mathematica notebook for each detector discussed under \verb!name.nb!, where \verb!name! is a place holder for the name of the detector. Currently,  notebooks are available for all of the detectors listed in Tab.~\ref{tab:cases}. Within each notebook, the geometric configurations of the detector are defined in terms of parametric equations of each surface element. The notebook reads in a list(s) of LLP particle data $(\theta, \phi, b)$, and outputs $\langle P((\text{LLP}) \text{ in d.r}) \rangle$ for a predetermined range of lifetimes. In the case of changes in detector designs, or new proposals, two general scripts were created for a box shaped and cylindrical shaped detectors; \verb!GeneralBox.nb! and \verb!GeneralCylinder.nb!. In both notebooks, the detector placement, dimensions, and rotation axis are defined by the user's specifications.  




\bibliographystyle{apsrev4-1}
\bibliography{refs}

\end{document}